# Surveying Wikipedians: a dataset of users and contributors' practices on Wikipedia in 8 languages

**Authors**

Caterina Cruciani[1], Léo Joubert[2], Nicolas Jullien[3], Laurent Mell[3], Sasha Piccione[1], Jeanne Vermeirsche[4]

**Affiliations**

[1] Department of Management, Università Ca'Foscari, Venice, Italy

[2] Normandie Univ, UNIROUEN, DYSOLAB, 76000 Rouen, France

[3] IMT Atlantique (Marsouin), Campus de Brest, CS 83818, Brest Cedex

[4] Université d'Avignon

**Corresponding author's email address and Twitter handle**

enquetes-wikipedia-2023@imt-atlantique.fr

**Abstract**

The dataset focuses on Wikipedia users and contains information about demographic and socioeconomic characteristics of the respondents and their activity on Wikipedia. The data was collected using a questionnaire available online between June and July 2023. The link to the questionnaire was distributed via a banner published in 8 languages on the Wikipedia page. Filling out the questionnaire was voluntary and not incentivised in any way. The survey includes 200 questions about: what people were doing on Wikipedia before clicking the link to the questionnaire; how they use Wikipedia as readers ("professional" and "personal" uses); their opinion on the quality, the thematic coverage, the importance of the encyclopaedia; the making of Wikipedia (how they think it is made, if they have ever contributed and how); their social, sport, artistic and cultural activities, both online and offline; their socio-economic characteristics including political beliefs, and trust propensities. More than 200 000 people opened the questionnaire, 102 862 started to answer, and constitute our dataset, and 10 704 finished it. Among other themes identified by future researchers, the dataset can be useful for advancing the research regarding the features of readers vs contributors of online commons, the relationship between trust, information, sources, and the use made of this information.

# SPECIFICATIONS TABLE

| | |
|---|---|
| **Subject** | Social science ; Political Science ; Sociology |
| **Specific subject area** | Use of a digital commons, Wikipedia: reading, contributing, access and use in the production of information, cultural activities |
| **Files joined to this paper** | *data_raw.csv* : the unfiltered database (102 862 respondents)<br>*data_filtered.csv* : the filtered database (10 704 respondents)<br>*codebook.xlsx* : codebook translated in 8 languages<br>*process.txt* : file giving an explanation of database construction process |
| **Data collection** | Data were collected online via a questionnaire developed with the open-source [limesurvey](#) software. A banner was shown to those opening Wikipedia in the different participating languages, asking to answer a questionnaire. The banner would redirect to a LimeSurvey page managed by the research centre Marsouin.org and hosted in France on OVHCloud servers.<br>The questionnaire (in English) is available here: https://questionnaires.marsouin.org/index.php/669835<br>The list of the languages in which the questionnaire is available can be found here: https://meta.wikimedia.org/wiki/Research:Surveying_readers_and_contributors_to_Wikipedia#Participating_languages<br>The questionnaires were identical across the languages, with the same coding for the questions and the answers, and are all provided alongside the data.<br>Different sources were used to build the questions, namely:<br>- the questions regarding the reader characteristics (reading habits) were taken from the questionnaire [done in 2019 by the WMF](#);<br>- the questions regarding reading behaviours and the assessment of the articles' readability benefited from the WMF's research "[Research:Understanding perception of readability in Wikipedia](#)"<br>- the questions regarding the contribution dimensions were taken from the Marsouin's [2015](#) survey<br>- the questions about the cultural dimensions were taken from the [e-FRAN IDEE research](#) and the ICT usages from the [ICT usage in households and by individuals European survey](#)<br>- the socio-demographic data were aimed at capturing the reading/contributing gaps [identified by the WMF](#) and are compatible with it (and with the European's and Marsouin's surveys), for the trust/values questions are taken from the [world value survey (wave 7)](#); |
| **Data source location** | The survey was carried out online; thus, the data were collected from everywhere. We collected the IP address of the device used to answer by the respondents. This personal information is not shared for privacy reasons. The dataset provides the city and the country where the device used was located (city & country calculated from the IP address using the R library: [rgeolocate](#)) |
| **Data accessibility** | Data identification number: 10.34847/nkl.4ecf4u8m<br>Direct URL to data: https://nakala.fr/10.34847/nkl.4ecf4u8m |

## VALUE OF THE DATA

The dataset provides information about the social and cultural characteristics of Wikipedia users and how they impact Wikipedia as a source of information. "Wikipedia" is today's encyclopaedia of reference. Its users and uses are highly varied (information, reference, documentary research, for training or more personal purposes, etc.)[1]. While there is a good understanding regarding its readers, how they use Wikipedia, and what acceptability as a source of information means, for them and for their "audience"[2] past studies did not delve very deeply into the social and cultural characteristics of the different users of Wikipedia.

The dataset offers insights regarding the perceived reliability and quality of Wikipedia content by its different users, addressing whether different degrees of generalised trust and different habits in terms of information map onto different evaluations. At the same time, this data might contribute to the literature on trust in the online environment, providing a focus on the nature of the information-production process within Wikipedia and its impact on Wikipedia as an information source. Currently, the literature on trust in social media distinguishes clearly between trust in the person providing the news and trust in the source outlet (see for example Sterret et al., 2019). By distinguishing between different types of Wikipedia users this dataset allows us to explore whether users and creators subscribe to different trust processes.

This dataset may help better understand how people start contributing to Wikipedia, why they keep doing it, and how the project could help them do so while fostering participation. Wikipedia is a volunteer-based organisation where individual contributions shape the offered contents sometimes above and beyond what readers might want to read about. The emerging misalignment between what people want to read and what the contributors are willing to provide (Warncke-Wang et al., 2015), alongside the decrease in the number of regular contributors in the main Wikipedia languages (Hill and Shaw, 2020) creates a tension between the quality of the information published and the possibility that contributions pursue individual positions and might be used to push one's own agenda. To address these constraints the Wikipedia platform has made contributing increasingly difficult, raising the barrier to contribution for the newcomers (ibid), but the jury is still out regarding whether this approach produces higher quality content. This database allows us to dwell more deeply in the motivations and the everyday practice of Wikipedians to understand contributions to the platform as a component of a more general habit of processing information and producing content in order to address in a more holistic way the issue of contribution quality.

Underlying motivations and the evolution of the profile of contributors are one of the main concerns of scholarly research since the rise of Wikipedia (Anthony et al., 2007). Given the complex and ever-evolving nature of Wikipedia, devising a theoretical and stable expectation about the evolution of contributors (not just in terms of numbers, but more importantly in terms of features and roles on the platform) is not easy. This database may help shed light on the relationship between some demographics and the likelihood of being and remaining a contributor. For instance, while we know that most wikipedians are male (Hill and Swah, 2013), we still need to establish how this specific feature interacts with roles taxonomy on the wiki or other biases in the coverage of wikipedia's corpus.

---

[1] On the history of Wikipedia, its impact and how it works, we recommend reading the book published by Reagle et Koerner (2020)
[2] see the survey of Wikipedia's readers in 14 languages by Lemmerich et al., 2019

Thanks to its set structure, which reduces any further effect of the design, structure and content of the website interface, Wikipedia is a very interesting set up over which the development of online trust can be studied. By reducing the drivers on online trust to content features, trust and credibility of Wikipedia can be assessed in a more controlled environment. This database allows to address how different patterns of information collection and use impact upon the credibility of the platform and correlate with more general dimensions of generalised trust.

## DATA DESCRIPTION

We provide two datasets, made available in the repository linked in the summary table:
- the *raw dataset*. It comprises 102 862 observations (amongst which only a half have really started to answer the questionnaire) obtained after removing duplicates identified by IP address matching (Sort rows so that the longest answers appear first. Filter duplicates by deleting individuals who answered all questions in the first part identically),
- the *filtered dataset*. It comprises 10 704 observations and is obtained from the raw dataset, by keeping only individuals that answered at least two questions after the age question. This is what we call a complete response.

We also built a .csv file with the code for each variable and the corresponding question (in 8 languages). These datasets are fully anonymized, for instance we do *not* provide wikipedia's nicknames even when the respondent gave it.

*Table 1. Summary of the diffusion process for each language.*

| Languages surveyed (alphabetic order) | Display dates for the banner demanding people to answer the questionnaire (all 2023) | Banner exposure regime (probability that it appears on the page opened. Page based, non (ip) session based | Number of people who clicked on the banner (opened the questionnaire) | Number of people who opened the questionnaire (went beyond the welcome page on the questionnaire site): our raw dataset | Number of complete responses: our filtered dataset |
|---|---|---|---|---|---|
| English | 06/02 -> 06/12 and 06/14 -> 06/17 | 5% | 70829 | 35555 | 4989 |
| French | 06/12 -> 06/26 | 30% | 22895 | 8129 | 815 |
| German | 06/28 -> 07/12 | 20% | 55948 | 21322 | 1993 |
| Italian | 06/17 -> 06/31 | 50% | 22077 | 9516 | 1092 |
| Portuguese | 06/27 -> 07/11 | 30% | 10162 | 3953 | 292 |
| Spanish | 06/27 -> 07/11 | 30% | 37302 | 18384 | 941 |
| Turkish | 06/01 -> 06/15 | 50% | 4657 | 1852 | 171 |
| Ukrainian | 06/14 -> 06/28 | 50% | 7002 | 4151 | 382 |
| **Total** | N/A | N/A | **230872** | **102862** | **10675** |

*Table 2. Detailed view of the filtering process*

| Languages surveyed (alphabetic order) | Number of people who opened the questionnaire (went beyond the welcome page on the questionnaire site): our raw dataset | Number of people who answered at least up to the question about what they were doing on Wikipedia before clicking the link to the questionnaire | Number of people who answered at least up to the question about how they use Wikipedia as a reader ("professional" and "personal" uses) | Number of people who answered at least up to the question about their valuation of Wikipedia | Number of people who answered at least up to up to the questions about whether they had ever contributed to Wikipedia | Number of people who answered at least up to up to the questions about their level of contribution | Number of people who answered at least up to their relationship to information/information trust | Number of people who answered at least up to their socio-economic characteristics | Number of complete responses: our filtered dataset |
|---|---|---|---|---|---|---|---|---|---|
| English | 35555 | 30664 | 16909 | 12491 | 9117 | 8306 | 6050 | 1880 | 4989 |
| French | 8129 | 5684 | 3366 | 1730 | 1272 | 1180 | 917 | 817 | 815 |
| German | 21322 | 16647 | 10233 | 5121 | 3817 | 3197 | 2171 | 1969 | 1993 |
| Italian | 9516 | 7763 | 4854 | 2338 | 1724 | 1552 | 1193 | 1125 | 1092 |
| Portuguese | 3953 | 3149 | 1743 | 791 | 562 | 448 | 313 | 292 | 292 |
| Spanish | 18384 | 15822 | 8200 | 3261 | 2205 | 1896 | 1119 | 967 | 941 |
| Turkish | 1852 | 1486 | 677 | 449 | 275 | 234 | 173 | 166 | 171 |
| Ukrainian | 4151 | 3947 | 2166 | 952 | 664 | 594 | 458 | 393 | 382 |
| **Total** | **102862** | **85162** | **48148** | **27133** | **19636** | **17407** | **12394** | **7609** | **10675** |

## EXPERIMENTAL DESIGN, MATERIALS AND METHODS

The dataset is the result of a survey carried out through LimeSurvey. The questionnaire was completed on a voluntary basis after clicking on a Wikipedia "banner" that would appear on the top of the Wikipedia page the users were browsing. The language of the questionnaire depended on the language in which the respondent opened Wikipedia (a banner in French targeting the questionnaire in French was published on the fr.wikipedia.org, in German for de.wikipedia.org, and so on) (See Table 1 and 2 for the complete list of languages). The questionnaire was available for a total of two weeks in each language. The overall data-collection process took place between the beginning of June 2023 and the middle of July 2023. Please refer to table 1 and 2 below for more information about the time-frame and events associated with the data collection and data curation.

The questionnaire was originally developed in French. It had been translated using [DeepL](#) and then reviewed by native speaking people (researchers or members of the Wikipedia community). The questionnaire was presented in the different languages' discussion spaces (in en.wikipedia: the Village Pump, in fr.wikipedia le bistro, and so on, but for one, as the community of de.wikipedia does not favour this type of annonce). The translations include local adaptation to make the questions more relevant to the different languages, such as the examples used for the scholar system to the local context, the questions regarding the language used (such as "is French your mother tongue" becoming in the Portuguese questionnaire "is Portuguese your mother tongue", etc.)

For each language but for the English we asked if the people had already read / contributed to the English Wikipedia, in addition to the Wikipedia they were using when they clicked on the banner.

## LIMITATIONS

Wikipedia is an ever growing community that spans across the globe; thus, it is difficult to identify the precise composition of this community. Given that the participation in this study was completely voluntary and not incentivised.

Moreover, self-selection bias might have played a role in participation. We cannot guarantee or even make the assumption that the collected sample is representative of the actual communities surveyed because we cannot grant that respondents and non-respondents are similar even given some set of characteristics.

Further, self selection and cultural features might have produced unequal response rates across the different languages. Regarding languages, their relative weight in the final sample is different and favour the English Wikipedia, but once again we cannot be certain that this over-representativeness is consistent with the actual population rates.

## ETHICS STATEMENT

Informed consent was obtained from participants and participant data has been fully anonymized

## CREDIT AUTHOR STATEMENT

Caterina Cruciani: question about trust in the media, translation in Italian

Léo Joubert: Conceptualization, Methodology, proofreading of Spanish translation

Nicolas Jullien: conceptualization, methodology, creation of the questionnaire, translation management, implementation on the platform (LimeSurvey), negotiation with the Wikimedia organisation to access the banners, translation in English

Laurent Mell: Conceptualization, Methodology, Formal analysis, Data Curation

Sasha Piccione: Formal analysis, Data Curation

Jeanne Vermeirsche: Conceptualization, Methodology, translation in English

# ACKNOWLEDGEMENTS

This research did not receive any specific grant from funding agencies in the public, commercial, or not-for-profit sectors.

It has received the help of several persons for the translation in this endeavour: User:Danielly Campos Dias for the Portuguese, User:Antanana for the Ukrainien, Müge Ozman for the Turkish, Anne Bartel-Radic re-read by utilisateur:Charlik for the German, François Élie who reread the French, Tania Jimenez for the Spanish (re-read by User:oscar-costero).

We benefited also from numerous feedback from the local communities before the official launch of the 2 week survey periods when we announced the questionnaire in the dedicated spaces of the village pumps.

We also received help from Anaïs Mazurier for the scrub data.

# DECLARATION OF COMPETING INTERESTS

The authors declare that they have no known competing financial interests or personal relationships that could have appeared to influence the work reported in this paper.

# REFERENCES

- Denise Anthony, Sean W. Smith, and Tim Williamson. 2007. *The quality of open source production: Zealots and good Samaritans in the case of Wikipedia*. Dartmouth College, Department of Computer Science.
- Benjamin Mako Hill and Aaron Shaw. 2013. The Wikipedia Gender Gap Revisited: Characterizing Survey Response Bias with Propensity Score Estimation. *PLoS ONE* 8, 6: e65782.
- Benjamin Mako Hill and Aaron Shaw. 2020. Wikipedia and the End of Open Collaboration. *Wikipedia* 20.
- David Sterrett, Dan Malato, Jennifer Benz, et al. 2019. Who Shared It?: Deciding What News to Trust on Social Media. *Digital Journalism* 7, 6: 783–801.
- Morten Warncke-Wang, Vivek Ranjan, Loren Terveen, and Brent Hecht. 2015. Misalignment Between Supply and Demand of Quality Content in Peer Production Communities. *Proceedings of the International AAAI Conference on Web and Social Media* 9, 1: 493–502.